\documentclass{ws-procs975x65}            
\usepackage{epsfig}
\begin{document}
\title{Thick brane solitons breaking $Z_2$ symmetry}

\author{Marzieh Peyravi$^{1}$, Nematollah
Riazi$^{2}$ and Francisco S. N.
Lobo$^{3}$}

\address{$^{1}$Department of Physics, School of Sciences, Ferdowsi University
of Mashhad, Mashhad 91775-1436, Iran\\
E-mail: marziyeh.peyravi@stu-mail.um.ac.ir\\}


\address{$^{2}$Physics Department, Shahid Beheshti University, Evin, Tehran 19839, Iran\\
E-mail: n\_riazi@sbu.ac.ir}


\address{$^{3}$Instituto de Astrof\'{\i}sica e Ci\^{e}ncias do Espa\c{c}o, Faculdade de
Ci\^encias da Universidade de Lisboa, Edif\'{\i}cio C8, Campo Grande,
P-1749-016 Lisbon, Portugal\\
E-mail: fslobo@fc.ul.pt}

\begin{abstract}
 New soliton solutions for thick branes in 4 + 1 dimensions are considered in this article. In particular, brane models based on the sine-Gordon (SG), $\varphi^{4}$ and $\varphi^{6}$  scalar fields are investigated; in some cases $Z_{2}$ symmetry is broken. Besides, these soliton solutions are responsible for supporting and stabilizing the thick branes. In these models, the origin of the symmetry breaking resides in the fact that the modified scalar field potential may have non-degenerate vacuua and these non-degenerate  vacuua determine the cosmological constant on both sides of the brane.  At last, in order to explore the particle motion in the neighborhood of the brane, the geodesic equations along the fifth dimension are studied.
\end{abstract}


\bodymatter

\section{Introduction}\label{aba:sec1}
Since there is no known fundamental principle requiring spacetime to be $(3 + 1)-$dimensional \cite{Ge2},  our observable universe might be a $(3+1)-$dimensional brane in a higher dimensional bulk \cite{lin,Ge3}. Most models suggest that there are one or more flat 3-branes embedded discontinuously in the ambient geometry \cite{de}. Furthermore, almost all of the extra-dimensional models require the existence of scalar fields, for instance, to generate a domain-wall which localizes matter fields \cite{Ge2}. Scalar fields serve to stabilize the size of the compact extra dimensions\cite{lin}, and can support the modification of the Randall-Sundrum warped-space \cite{Randall:1999ee,Randall:1999vf} to a smoothed-out version \cite{Ge3,de}, or to cut off the extra dimension at a singularity \cite{CG,Ge1}.
In addition to this, the nonlinearity in the scalar field and, in particular, the existence of disconnected vacuua in the self-interaction of the scalar field lead to the appearance of a stable localized solution, which is a good motivation for building thick brane models \cite{blg}. It is frequently invoked to replace an infinitely thin brane with a thick one, by supposing a scalar field with soliton behavior.
Based on this point of view, in this paper we investigate diverse models, namely, the sine-Gordon (SG), $\varphi^{4}$ and $\varphi^{6}$ brane models which have broken $Z_{2}$ symmetry in some cases. Note that the $Z_{2}$ symmetry may be restored by a proper choice of model parameters. The origin of symmetry breaking in our models reside in the fact that the modified scalar field potential may have non-degenerate vacuua \cite{Peyravi:2015bra}. The cosmological constant on both sides of the brane are determine by these vacuua.

We consider a thick brane, embedded in a five-dimensional (5D) bulk spacetime, modelled
by the following action:
\begin{equation}
S=\int
d^{5}x\sqrt{|g|}\left[\frac{1}{4}R-\frac{1}{2}\partial_{\alpha}\varphi\partial^{\alpha}\varphi-V(\varphi)\right].
\end{equation}
 The simplest line element of the brane, embedded in the  five-dimensional bulk space-time
can be written as\cite{br}:
\begin{eqnarray}
ds^{2}_{5}&=&g_{AB}dx^{A}dx^{B}\nonumber\\
&=&dw^{2}+e^{2A}(dx^{2}+dy^{2}+dz^{2}-dt^{2}),
\end{eqnarray}
where $A$ is the warp factor which depends only on the 5D coordinate $w$.

The 5D gravitational and scalar field equations take respectively the following forms
\begin{eqnarray}
3A^{\prime\prime}+6{A^{\prime}}^{2}&=&-\kappa_{5}^{2}e^{-2A}T_{00}=-\kappa_{5}^{2}\left[\frac{1}{2}{\varphi^{\prime}}^{2}+V(\varphi)\right] ,
     \\
6{A^{\prime}}^{2}&=&\kappa_{5}^{2}T_{44}=\kappa_{5}^{2}\left[\frac{1}{2}{\varphi^{\prime}}^{2}-V(\varphi)\right] ,
     \\
\varphi^{\prime\prime}+4A^{\prime}\varphi^{\prime}&=&\frac{d
V(\varphi)}{d\varphi},
\end{eqnarray}
where the prime denotes  derivative with respect to $w$.

In order to obtain the first-order equation, one can introduce the
auxiliary function $W$ in such a way that:
\begin{eqnarray}
A^{\prime}&=&-\frac{1}{3}W(\varphi), \nonumber\\
\varphi^{\prime}&=&\frac{1}{2}\frac{\partial
W(\varphi)}{\partial\varphi},
\end{eqnarray}
while $V(\varphi)$ achieves the special form \cite{de,Sa,Af,CM}:
\begin{equation}
V(\varphi)=\frac{1}{8}\left(\frac{\partial
W(\varphi)}{\partial\varphi}\right)^{2}-\frac{1}{3}W(\varphi)^{2}.
\end{equation}
Moreover, it may be instructive to calculate the geodesic equation along the fifth dimension in a thick brane, in order to investigate the particle motion near the brane \cite{JS}. To this end, we start with the geodesic equation:
\begin{eqnarray}
\frac{d^{2}x^{0}}{d\tau^{2}}+\Gamma^{0}_{AB}\frac{dx^{A}}{d\tau}\frac{dx^{B}}{d\tau}&=&0   \qquad
\Rightarrow  \qquad  \frac{d}{d\tau}\left(-2e^{2A}\dot{t}\right)=0,  \nonumber\\
\frac{d^{2}x^{4}}{d\tau^{2}}+\Gamma^{4}_{AB}\frac{dx^{A}}{d\tau}\frac{dx^{B}}{d\tau}&=&0 \qquad
\Rightarrow \qquad \ddot{w}+A'e^{2A}\dot{t}^{2}=0 ,
\end{eqnarray}
which leads to
\textbf{\begin{equation}\label{geo}
\ddot{w}+c_{1}^{2}f(w)=0 ,
\end{equation}}
where $c_{1}$ is a constant of integration and the function $f(w)$ is defined as
\begin{equation}
f(w)=A'(w)e^{-2A(w)}.
\end{equation}

\section{Thick brane models}
\begin{figure}
\epsfxsize=13.2cm\centerline{\epsfbox{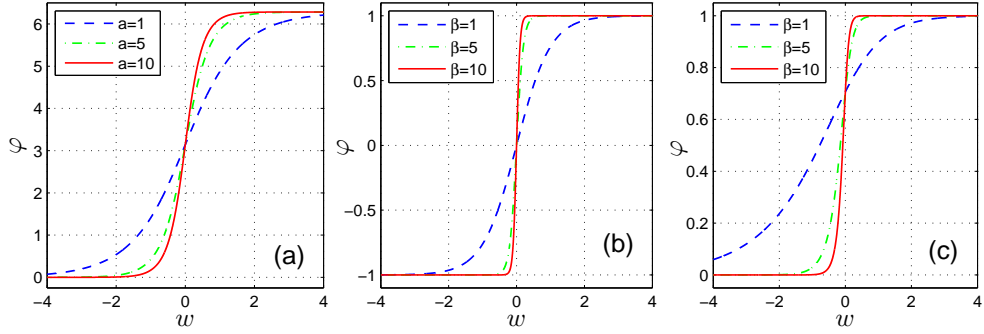}} \caption{Soliton
solutions as a function of the fifth dimension for the following models: (a) SG for $b=1$, (b) $\phi^4$ for $\alpha=1$ and (c) $\phi^6$ for $\alpha=1$ systems.
Dashed, dotted-dashed, and continuous curves correspond to solitons
with decreasing brane thickness. In the limit of an infinite $a/\beta$
parameter, the soliton approaches the step function.\label{st}}
\end{figure}

\begin{figure}
\epsfxsize=13.6cm\centerline{\epsfbox{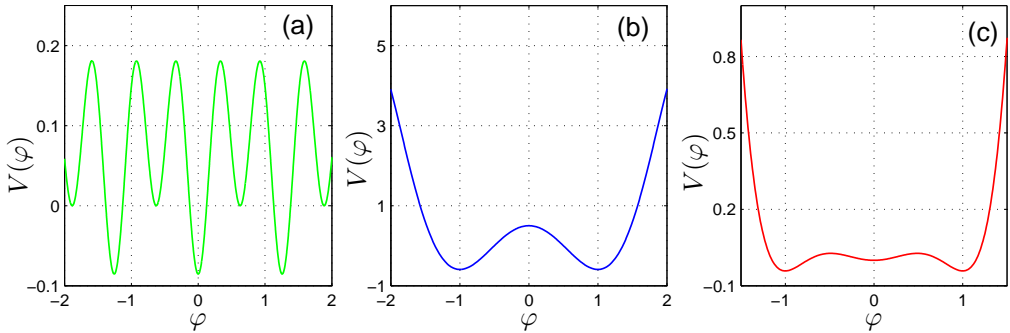}} \caption{The plots depict
the modified soliton potential  as a function of the fifth dimension
for (a) SG with $a=b=10$, (b) $\phi^4$ with $\alpha=\beta=1$ and (c) $\phi^6$ with
$\alpha=\beta=1$ systems.  The potential of the SG and $\phi^6$
systems have non-degenerate vacuua. In contrast, the $\phi^4$
potential has degenerate vacuua and this leads to stable,
topological solitonic brane.\label{v}}
\end{figure}

First we consider the Sine-Gordon (SG) system, where the self-interaction potential of this model reads:
\begin{equation}
\tilde{V}(\varphi)=\frac{a}{b}\left[1-cos(b\varphi)\right]\,,
\end{equation}
where $a$ and $b$ are free parameters of the model.
The SG system has the following exact static kink solution \cite{Ri}:
\begin{equation}
\varphi(w)=\frac{4}{b}\arctan\left(e^{\sqrt{ab}w}\right),
  \label{SGdilaton}
\end{equation}
which is plotted in Fig.\ref{st}(a), for various values of parameters $a$ and $b$, which correspond to branes with different thickness.
When considered as the brane potential, however, this potential
should be modified to become consistent with the Einstein equations.
The corresponding potential for this model is given by:
\begin{equation}
V(\varphi)=\frac{2a}{b}\sin^{2}\left(\frac{b\varphi}{2}\right)-\frac{64a}{3b^{3}}\left[1+\cos\left(\frac{b\varphi}{2}\right)\right]^{2}\,,
  \label{SGpotential}
\end{equation}
which is depicted in Fig.\ref{v}(a). Notice that this potential has two series of  non-degenerate vacuua, as in the DSG  (double sine-Gordon) system
potential \cite{Pey}. However, in the limit of $b\gg a$ these vacuua
tend to the same value (become degenerate), such as the potentials used in \cite{blg,CM}.

\begin{figure}
\epsfxsize=13.2cm\centerline{\epsfbox{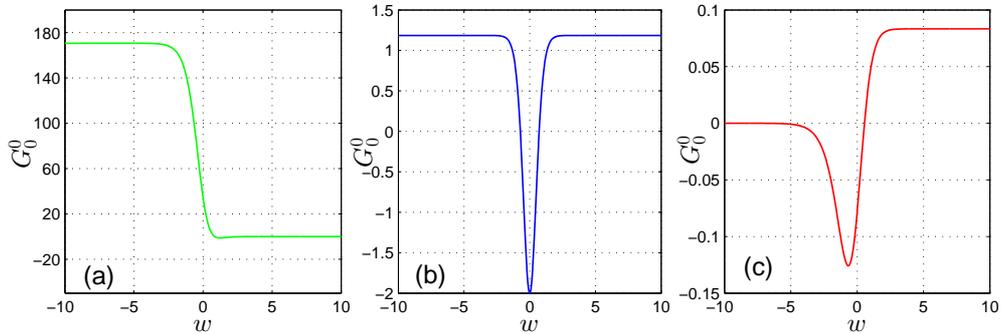}} \caption{The plots
depict the Einstein tensor component $G_{0}^{0}$ for the (a) SG with
$a=b=1$, (b) $\phi^4$ with $\alpha=\beta=1$ and (c) $\phi^6$ with
$\alpha=\beta=1$ systems. Note that this quantity approaches
different constant values for the SG and $\varphi^{6}$ systems,
while the $\varphi^{4}$ system is $Z_{2}$-symmetric.\label{gw}}
\end{figure}

In this case the energy density is localized
at the brane and the thickness of the latter is $\triangle=\frac{1}{2\sqrt{ab}}.$
It can be seen that there is no singularity in the Ricci scalar
and/or Kretschmann scalar. Moreover, in the limits of $w\rightarrow\pm\infty$, the Ricci scalar and  all the components of the Einstein in the right and left sides of the brane  are different (Fig.\ref{gw}(a)).
However, in the limit of $w\rightarrow0$, the Einstein tensor is given by $G^{\mu}_{\nu}=\frac{8}{3}\frac{(16-3b^{2})a}{b^{3}}\,\delta^{\mu}_{\nu}$ (for $\mu=\nu=0,1,2,3$),
and one can interpret it as the cosmological constant on the brane, i.e.,
$G^{i}_{j}\propto\Lambda\delta^{i}_{j}$.

Therefore, we have a broken $Z_{2}$-symmetry in the bulk, as the two sides of the brane differ completely. On the right and in the limit of $w\rightarrow+\infty$, the Einstein tensor and consequently the cosmological constant of the bulk vanish, so the bulk is asymptotically Minkowski. However, on the other side of the brane, these quantities are nonzero and equal to the constant value
$512 a/(3b^{3})$, and as a result the bulk would be
de Sitter. Besides, by checking the geodesic equation of a test particle moving only in
the direction of the extra dimension, the confining effect of the scalar field is proved.

The second brane model is the $\varphi^{4}$-based model, for which we have \cite{blg}:
\begin{equation}
\tilde{V}(\varphi)=\frac{\beta^{2}}{2\alpha^{2}}\left(\varphi^{2}-\alpha^{2}\right)^{2},
\end{equation}
where $\alpha$ and $\beta$ are constants. The kink solution reads
\begin{equation}
\varphi(w)=\alpha\tanh(\beta w),
\end{equation}
which is depicted in Fig.\ref{st}(b).
The potential is obtained as
\begin{equation}
V(\varphi)=\frac{1}{2}\alpha^{2}\beta^{2}\left(1-\frac{\varphi^{2}}{\alpha^{2}}\right)^{2}-\frac{4}{27}\varphi^{2}\alpha^{2}\beta^{2}\left(3-\frac{\varphi^{2}}{\alpha^{2}}\right)^{2}\,,
\end{equation}
which is plotted in Fig.\ref{v}(b) and the brane thickness becomes $\triangle=\beta^{-1}$.
As for the previous SG model, we determine the limits $w\rightarrow\pm\infty$ for all the components of the Einstein tensor components, which are the same at different sides and given by $\frac{32}{27}\alpha^{4}\beta^{2}$ (Fig.\ref{gw}(b))
and in the limit of $w\rightarrow0$ the Einstein tensor components (for
$\mu=\nu=0,1,2,3$) takes the form $-2\alpha^{2}\beta^{2}\delta^{\mu}_{\nu}$.
Therefore, in this model the cosmological constant on the brane $\Lambda$ would be $-2\alpha^{2}\beta^{2}$.
For this model a test particle oscillate around the brane by $\Omega=\sqrt{\frac{2}{3}}c_{1}\alpha\beta$.

The last model is based on the $\varphi^{6}$ soliton. For this model, we have the following potential:
\begin{equation}
\tilde{V}(\varphi)=\frac{\beta^{2}}{4\alpha^{2}}\varphi^{2}\left(\varphi^{2}-\alpha^{2}\right)^{2}\,,
\end{equation}
where $\alpha$ and $\beta$ are constant(as in the
$\varphi^{4}$ model) and as a result, the kink solution is given by \cite{hosein}:
\begin{equation}
\phi (w)=\frac{\alpha}{\sqrt{1+e^{(-\sqrt{2}\alpha\beta w)}}}.
\end{equation}
The kink solution is depicted in Fig.\ref{st}(c).
Thus, for the $\varphi^{6}$ system the self-interaction potential takes the form (Fig.\ref{v}(c)):
\begin{equation}
V(\varphi)=\frac{1}{4}\frac{\beta^{2}\varphi^{2}\left(\alpha^{2}-\varphi^{2}\right)^{2}}{\alpha^{2}}-\frac{1}{24}\frac{\beta^{2}\varphi^{4}\left(2\alpha^{2}-\varphi^{2}\right)^{2}}{\alpha^{2}}\,,
\end{equation}
The thickness of this brane is given by $\triangle=(\sqrt{2}\alpha\beta)^{-1}$. Moreover, for this model and in the limit of
$w\longrightarrow\pm\infty$ the mixed Einstein tensor components reduce to $\frac{1}{12}\alpha^{6}\beta^{2}$ and zero respectively see Fig.\ref{gw}(c).
and in the limit of $w\rightarrow0$, the Einstein tensor is $G^{\mu}_{\nu}= \frac{1}{8}\alpha^{4}\beta^{2} \left(\frac{3}{8}\alpha^{2}-1\right)\, \delta^{\mu}_{\nu}$.
\section{Conclusion}

In this work, we obtained exact thick brane models inspired by
well-known nonlinear systems, namely, the sine-Gordon ($SG$),
$\varphi^{4}$ and $\varphi^{6}$ models. The confining effect of the
scalar field in all these three models were confirmed by examining
the geodesic equation for a test particle moving normal to the
brane. In particular, it turns out that the modified potential for
the $SG$ system resembles that of the double sine-Gordon ($DSG$)
system, while those of $\varphi^{4}$ and $\varphi^{6}$ became
$\varphi^{6}$ and $\varphi^{8}$, respectively.

In the case of the $SG$ model,
the resulting brane does not have $Z_2$ symmetry, in general, where
the center of the brane may be displaced from $w=0$ and the
potential will not be an odd function of $w$ in general. However, by
a suitable choice of the model parameters it is possible to make the
vacuua of the effective potential degenerate, in which case the
$Z_2$ symmetry is restored. In the case of the $\phi^6$ model,
however, we could not restore this symmetry via re-parametrization.

\section{Acknowledgments}
M.P. acknowledges the support of
Ferdowsi University of Mashhad via the proposal No. 32361. N.R. acknowledges the support of Shahid Beheshti
University Research Council. F.S.N.L.
acknowledges financial  support of the Funda\c{c}\~{a}o para a
Ci\^{e}ncia e Tecnologia through an Investigador FCT Research
contract, with reference IF/00859/2012, funded by FCT/MCTES
(Portugal), and the grant EXPL/FIS-AST/1608/2013.



\end{document}